\documentclass[10pt]{article}
\usepackage{graphicx}
\usepackage[T1]{fontenc}
\usepackage[bottom=2cm,top=3cm,left=3cm,right=2cm]{geometry}
\begin{document}
\title{Collisionless self-gravitating systems in $f(R)$-gravity within Palatini approach and relativistic Boltzmann equation in the Newtonian approach}

\author{Ra\'{i}la~Andr\'e and Gilberto~M~Kremer \\
Departamento de F\'{i}sica, Universidade Federal do Paran\'a, Curitiba, Brazi}


\date{\today}
\maketitle

\begin{abstract}
In this work we analyze the dynamics of collisionless self-gravitating systems described by the $f(R)$-gravity and Boltzmann equation in the weak field approximation, focusing on the Jeans instability for theses systems. The field equations in this approximation were obtained within the Palatini formalism. Through the solution of coupled equations we achieved the collapse criterion for infinite homogeneous fluid and stellar systems, which is given by a dispersion relation. This result is compared with the results of the standard case and the case for $f(R)$-gravity in metric formalism, in order to see the difference among them. The limit of instability varies according to which theory of gravity is adopted.
\end{abstract}

\section{Introduction}

The General Relativity (GR) has emerged as a highly successful theory for  cosmological models, surviving various tests.  In search of a generalization of GR emerged a theory of gravity known as modified $f(R)$-gravity. An important consequence of this new theory resides in the fact that it is no longer necessary to introduce unknown entities to explain the accelerated expansion of the Universe and the formation of structures. Searching for further generalization, it was introduced also the Palatini formalism where the most fundamental aspect to be noted  is the independence, \textit{a priori}, between the metric tensor and affine connection.

In this work, in order to study the formation of structures, we adopt a model described by $f(R)$-gravity from the point of view of the Palatini approach, which is able to describe the dynamics and the collapse of collisionless self-gravitating systems. To investigate this dynamics, it becomes necessary the introduction of the collisionless Boltzmann and Poisson equations for the fields. The Boltzmann equation is an essential tool for the understanding of the processes that occurs in interstellar clouds, as the damping waves perturbations.
Here we investigated the Jeans instability for physical systems that exhibit weak gravitational field, with slow variation or static and  objects which moves slowly compared to the speed of light. This instability causes the collapse of interstellar gas clouds and hence, the formation of structures. This occurs when the internal pressure of the gas is not enough to prevent gravitational collapse.
The aim of this work is to show the difference in the behavior between the solutions obtained from this model and from those obtained by $f(R)$-gravity in the metric formalism in the Newtonian limit [1, 2] and by Newtonian gravity [3]. Here we adopt the signature $(-, +,+,+)$ and follow the book [4] for the conventions of the Riemann tensor and its contractions.

\section{Palatini formalism for $f(R)$-gravity in the Newtonian limit}
\label{sec:one}

The action which represent the modified gravity theories reads
\begin{eqnarray} \label{1}
S= \int dx^{4} \sqrt{-g} \left[ -f(R)/(2\kappa) +  \mathcal{L}_m \right]
\end{eqnarray}
where $f(R)$ is an analytic function of the Ricci scalar $R$, $\kappa={8 \pi G}/{c^{4}}$  denotes the  gravitational field coupling and $\mathcal{L}_m$ is the Lagrangian density of the matter field.

From the variation of the action (\ref{1}) according to the Palatini formalism we get the modified Einstein's field equations
\begin{eqnarray} \label{2}
f'(R) R_{\mu \nu}-\frac{1}{2}g_{\mu \nu} f(R)=- \kappa T_{\mu \nu}=\frac{2\kappa}{\sqrt{-g}}\frac{\delta\left(\mathcal{L}_m \sqrt{-g}\right)}{\delta g^{\mu\nu}}.
\end{eqnarray}
Above $f'\equiv f'(R)= df(R)/dR$ and $T_{\mu\nu}$ is the energy-momentum tensor of the gravitational sources. The Ricci tensor $R_{\mu\nu}$ and the affine connection ${\Gamma}^{\alpha}_{\mu \nu}$ are given in terms of the Riemannian Ricci tensor $\widetilde R_{\mu\nu}$ and connection $\widetilde{\Gamma}^{\alpha}_{\mu \nu}$ by
\begin{eqnarray} \label{3a}
R_{\mu\nu}=\widetilde R_{\mu\nu}-\frac{3}{2f^{\prime2}}\partial_\mu f'\partial _\nu f'+\frac{1}{f'}\widetilde\nabla_\mu \widetilde\nabla_\nu f'+\frac{1}{2f'}\widetilde\nabla^\sigma \widetilde\nabla_\sigma f' g_{\mu\nu},\\\label{3b}
\Gamma^{\alpha}_{\mu \nu}=\widetilde{\Gamma}^{\alpha}_{\mu \nu} + \frac{1}{2 f'} g^{\lambda \alpha} \left[ \partial_\mu f' g_{\lambda \nu} + \partial_{\nu} f' g_{\mu \lambda} - \partial_\lambda f' g_{\mu \nu}\right].
\end{eqnarray}

In the Newtonian approach the metric tensor can be written in terms of the Minkowski tensor $\eta_{\mu \nu}$ plus corrections of order $\overline{v}^2=\overline M G/\overline r$ [4] as $g_{00}=-1-g_{00}^{(2)}$ and $g_{ij}=\delta_{ij}+g_{ij}^{(2)}$. Up to this order we have
\begin{eqnarray} \label{4}
\widetilde R^{(2)}=-\frac12\nabla^2 g_{00}^{(2)}+\frac12\nabla^2 g_{ii}^{(2)}=\frac1{c^2}\nabla^2\left(\phi-\varphi\right),
\end{eqnarray}
where $\phi$ and $\varphi$ are  gravitational potentials associated with $g_{00}^{(2)}$ and $g_{ii}^{(2)}$, respectively.

In this work we consider  the following expression for $f(R)=R+f_2R^2$, where $f_2$ is a small quantity. The function proportional to $R^2$ is not able to reproduce an accelerated expansion of the Universe as it is currently observed. However, it should not be a problem, since the formation of structures, which is the focus of this work, occurred at a period much earlier than the current acceleration of the Universe.
In this case the trace of (\ref{3a}) in the $\overline{v}^2$ approximation reads
\begin{eqnarray} \left(1-6f_2\nabla^2\right)R^{(2)}=\widetilde R^{(2)},\end{eqnarray}
or by considering  $\left(1-6f_2\nabla^2\right)$ an invertible operator
\begin{eqnarray}\label{5} 
 R^{(2)}&\approx&\left(1+6f_2\nabla^2+36f_2^2\nabla^4\right)\widetilde R^{(2)} = \frac1{c^2}\left[\nabla^2\left(\phi-\varphi\right)+6f_2\nabla^4\left(\phi-\varphi\right)
 +36f_2^2\nabla^6\left(\phi-\varphi\right)\right],
\end{eqnarray}
which is an approximation up to the order $f_2^2$. Since
\begin{eqnarray} \widetilde R_{00}^{(2)}=\frac{\nabla^2 g_{00}^{(2)}}{2}=-\frac{\nabla^2\phi}{c^2},
\end{eqnarray}
the time component of the Ricci tensor  (\ref{3a}) in the $\overline{v}^2$ approximation reduces to
\begin{eqnarray} \label{6}
R_{00}^{(2)}=\widetilde R_{00}^{(2)}-f_2\nabla^2 R^{(2)}\approx-\frac1{c^2}\left[\nabla^2\phi+f_2\nabla^4\left(\phi-\varphi\right)
 +6f_2^2\nabla^6\left(\phi-\varphi\right)\right].
\end{eqnarray}

The source of the gravitational field is a pressureless fluid where the components of the  energy-momentum tensor are given by $T_{\mu\nu}=(\rho c^2,0,0,0)$ with $\rho$ denoting the fluid mass density.

Now the time component and the trace of Einstein's field equations (\ref{2}) together with (\ref{5}) and (\ref{6}) leads to the following system of Poisson equations
\begin{eqnarray} \label{7a}
\nabla^{2} (\phi+\varphi)- 4 f_2 \nabla^{4}(\phi -\varphi)-24 f_2^2 \nabla^6(\phi-\varphi)= 16 \pi G \rho,
\\ \label{7b}
\nabla^{2}(\phi-\varphi)+ 6 f_2 \nabla^{4}(\phi-\varphi)+36 f_2^2\nabla^6(\phi-\varphi)= -8 \pi G \rho,
\end{eqnarray}
respectively.
If we sum the above equations and consider $f_2=0$, the standard Poisson equation is recovered.

\section{Jeans instability in the framework of Boltzmann equation}

Now the aim is to obtain from the collisionless Boltzmann equation in the Newtonian limit a criterion for the  collapse of stellar systems, which is related with a dispersion relation. The collisionless Boltzmann equation has the following form in the Newtonian limit
\begin{eqnarray} \label{8}
\frac{\partial f}{\partial t}+ (\vec{v}\cdot \vec{\nabla}_r)f-(\vec{\nabla}\phi \cdot \vec{\nabla}_v)f=0,
\end{eqnarray}
where $f \equiv f(\vec{r},\vec{v},t) $ is the distribution function, which gives the mass density of the stellar system through
$\rho(\vec{r}, t)=\int f(\vec{r},\vec{v},t) d\vec{v}.$

We consider that the self-gravitating equilibrium system -- described by a time-independent distribution function $f_{0}(\vec{r},\vec{v})$ and potentials $\phi_{0}(\vec{r})$ and $\varphi_{0}(\vec{r})$ -- is subjected to a small perturbation, namely,
$f(\vec{r},\vec{v},t)=f_{0}(\vec{r},\vec{v})+\epsilon f_{1}(\vec{r},\vec{v},t)$,
$\phi(\vec{r},t)=\phi_{0}(\vec{r})+\epsilon \phi_{1}(\vec{r},t)$ and $\varphi(\vec{r},t)=\varphi_{0}(\vec{r})+\epsilon \varphi_{1}(\vec{r},t)$, where $\epsilon \ll 1$. The equilibrium for a homogeneous system is achieved by Jeans "swindle" that allows us to make $\phi_0=0$ and $\varphi_0=0$ without loss of consistency. After these considerations, we linearize the Boltzmann (\ref{8}) and the field equations (\ref{7a}), (\ref{7b})   through the substitution of the above conditions.  Subsequently we write the resulting equations in Fourier space as follows
\begin{eqnarray} \label{10a}
-i \omega \overline f_1+\vec{v} \cdot (i \vec{k}\,\overline f_1)-(i \vec{k}\, \overline\phi_1) \cdot \frac{\partial \overline f_0}{\partial \vec{v}}=0,
\\\label{10b}
-k^{2}(\overline\phi_1+\overline\varphi_1)-4f_2 k^{4}(\overline\phi_1 - \overline\varphi_1)+ 24 f_2^2 k^6(\overline\phi_1-\overline\varphi_1)= 16 \pi G \int \overline f_1 d\vec{v},
\\ \label{10c}
k^{2}(\overline\phi_{1}-\overline\varphi_{1})-6f_2 k^{4}(\overline\phi_1 -\overline\varphi_1)+36 f_2^2 k^6(\overline\phi_1-\overline\varphi_1)= 8 \pi G \int \overline f_1 d\vec{v},
\end{eqnarray}
where the overbarred quantities indicate the Fourier transforms in the $(\omega,\vec{k})$ space.

By eliminating the overbarred quantities from the system of equations (\ref{10a}) - (\ref{10c})  it follows the dispersion relation
\begin{eqnarray} \label{11}
1+ \frac{4 \pi G}{k^{2}}\frac{(1-8 f_2k^2+48 f_2^2k^4)}{(1-6f_2k^2+ 36 f_2^2 k^4)}\int \left(\frac{\vec{k}\cdot \frac{\partial f_0}{\partial \vec{v}}}{\vec{v}\cdot \vec{k}-\omega} \right) d\vec{v}=0.
\end{eqnarray}

In stellar systems one assumes usually the Maxwell distribution function
\begin{eqnarray} \label{12}
f_0(\vec{v})=\frac{\rho_0}{ (2\pi\sigma^2)^\frac32 }e^{-\frac{v^2}{2\sigma^2}},
\end{eqnarray}
where $\sigma$ is  a dispersion  velocity and $\rho_0$  a constant mass density.

Without loss of generality we can choose $\vec{k}=(k,0,0)$ so that the dispersion relation  (\ref{11}) together with (\ref{12}) can be integrated with respect to the velocity components $v_y$ and $v_z$, yielding
\begin{eqnarray} \label{13}
\frac{k^2}{k_J^2}-\frac{(1-8 f_2k^2+ 48 f_2^2k^4)}{(1-6f_2k^2+ 36 f_2^2 k^4)}\frac{2}{\sqrt{\pi}} \int_{0}^\infty \frac{x^2 e^{-x^2}}{x^2 -\omega^2/(2\sigma^2 k^2)}dx=0.
\end{eqnarray}
Here we have introduced the Jeans wavenumber $k_J=\sqrt{4\pi G\rho_0/\sigma^2}$ and the integration variable $x=v_x/(\sqrt2 \,\sigma)$.

Unstable solutions are such that $\Re( {\omega})=0$ and $\omega_I=\Im(\omega)>0$ [3]. In this case the integral on the right-hand side of (\ref{13}) can be evaluated [5] and the dispersion relation (\ref{13}) reduces to
\begin{eqnarray} \label{14}
\frac{\left\langle9 \frac{k^6}{k^{6}_J}\right\rangle+\left(3\frac{k^4}{k^4_J}\right)+\frac{k^2}{k^{2}_J}}{\left\langle12 \frac{k^4}{k^4_J}\right\rangle+\left(4\frac{k^2}{k^2_J}\right)+1 }=\left\{1-  \frac{\sqrt{\pi}\omega_I}{\sqrt{8\pi G\rho_0} }\frac{k_J}{k}\, e^{\left(\frac{\omega_I^2}{8\pi G\rho_0}\frac{k_J^2}{k^2}\right)}{\rm erfc}\left(\frac{\omega_I}{\sqrt{8\pi G\rho_0} }\frac{k_J}{k}\right)\right \}.
\end{eqnarray}
Above ${\rm erfc}$ is the complementary error function and we have introduced the same parametrization for $f_2=-1/(2k_J^2)$ that was adopted by [1]. The choice of this parameterization was made with the intention of describing a dimensionless ratio and to compare directly with the work done in [1]. Other choices could have been made, but they all culminate in the same result, differing only by constants. Without the terms within the brackets in the numerator and denominator on the left-hand side of (\ref{14}), it follows the classical dispersion relation for Newtonian gravity. The terms within round brackets refer to the contribution of the $f(R)$ theory in the metric formalism [1], while the ones within the round and angular brackets correspond to the Palatini formalism.

The solution of (\ref{14}) for $\omega_I=0$ furnishes the following limiting values for the wavenumber: (a) $k^2=1.3171\,k_J^2$ for the $f(R)$ theory in the Palatini formalism, (b) $k^2=1.2638\, k_J^2$ for the $f(R)$ theory in the metric formalism and (c) $k^2=k_J^2$ in the Newtonian theory.
For comparison, the dispersion relation  (\ref{14}) is plotted in Fig. 1  together with the results obtained from the $f(R)$ theory in the metric formalism and from the Newtonian gravity. The gravitational collapse occurs for the unstable wave solutions where the values of  $k^2/k_J^2$ lie below the curves in Fig. 1. Above these values the oscillations remain stable. We infer from this figure that the $f(R)$ theory in the Palatini formalism furnishes a large range of solutions than the metric one, which in its turn is larger than the one of the Newtonian theory.

The stellar system is stable up to a critical mass and if this critical value is exceeded the gravitational collapse occurs. In the Newtonian theory  this critical value is the Jeans mass $M_J$, which is defined as the mass within a sphere of diameter $\lambda_J=2\pi/k_J$, i.e, $M_J=4\pi\rho_0(\pi/k_J)^3/3$. From the above results  one obtain the following ratios: (a) $M/M_J=0.662$ for the $f(R)$ theory  in the Palatini formalism and (b) $M/M_J=0.704$ for the $f(R)$ theory  in the metric formalism. Hence the $f(R)$ theory  in the Palatini formalism furnishes the smallest critical mass for the occurrence of the gravitational collapse.

\begin{figure}
\begin{center}
\includegraphics[width=6.9cm]{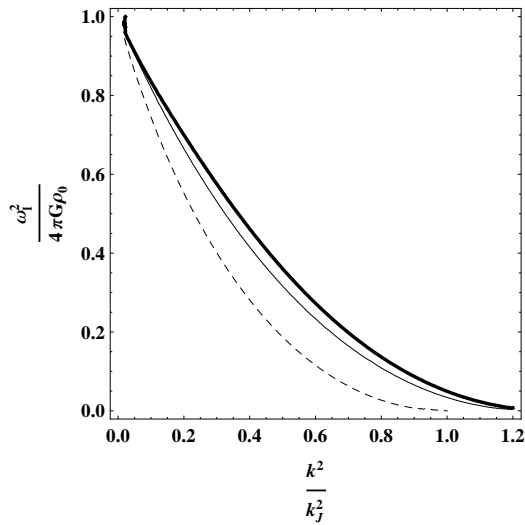}
\caption{\small  Ratio of wavenumbers $k^2/k_J^2$ versus normalized frequency $\omega_I^2/(4\pi G\rho_0)$ from the dispersion relation (\ref{14})  for $f(R)$-gravity in Palatini formalism (bold line), for $f(R)$-gravity in metric formalism (thin line) and for Newtonian gravity  (dashed line).}
\end{center}
\end{figure}

\section{Conclusions}

The dispersion relation is a collapse criterion   for infinite homogeneous fluid and stellar systems. Here, this relation is used to study the instability of collisionless systems. Figure 1 shows the behavior of the dispersion relation of  the models: Newtonian gravity, $f(R)$-gravity in the metric formulation and this one in the Palatini formalism. It is worth stressing that the characteristic wavenumber is given in terms of the classical one. The model proved to have a higher instability limit when compared to the others. The collapse occurs for higher  values of the wavenumber when compared with the Jeans wavenumber, so that critical mass of interstellar clouds decreases, modifying the initial conditions to start the collapse. The dispersion relation obtained via Palatini formulation differs from that obtained via metric formulation only for terms of order up to $f_2^2$; below this order, no difference was found. The terms of order $f_2^2$ arise only in the Palatini solution, when one writes the generalized Ricci scalar depending on the usual one. Thus, we can conclude that the differences reported here are a direct result of using the Palatini formalism.

The Jeans mass is a criterion for the study of gravitational collapse of interstellar clouds. The cloud is stable for sufficiently small mass (at a given temperature and radius), smaller than the limit established by Jeans mass. If this cloud exceeds  this mass limit, it starts a process of contraction until some other force can impede the total collapse. It was shown that the Jeans mass for the systems described by $f(R)$-gravity via Palatini formalism exhibits values smaller than the ones found for the metric case, and especially for Newtonian gravitation. This demonstrates that the limit for initiating the collapse of an interstellar cloud is below the classical limit favoring more the formation of structures. We may interpret this result by saying that  the same amount of matter is able to produce a larger curvature, causing a larger acceleration of test particles. Hence the test particles  describe a different geodesic than that they would follow in the metric formalism. In this way, we conclude that in the Palatini formalism of $f(R)$ gravity the formation of structures is more efficient than in the metric formalism of $f(R)$ gravity and Newtonian gravity.
\vspace{0.2cm}

{\bf Acknowledgments:} GMK acknowledges the support from the Conselho Nacional de Desenvolvimento Cient\' {i}fico e Tecnol\' {o}gico (CNPq), Brazil.
\vspace{0.3cm}
\newline
\noindent\textbf{References}
\newline
[1] S. Capozziello et al., Physical Review D {\bf 85} 044022 (2012).
\newline
[2] S. Capozziello and M. De Laurentis, Physics Reports {\bf 509} 167-321 (2011).
\newline
[3] J. Binney and S. Tremaine, \emph{Galatic Dynamics} (Princeton University Press, 2008).
\newline
[4] S. Weinberg, \emph{Gravitation and Cosmology, principles and applications of the general theory of relativity} (Wiley \& Sons, 1972).
\newline
[5] I. Gradshteyn and I. Ryzhiz, \emph{Table of Integrals, Series and Products} (Academic Press, 2007).




\end{document}